\documentclass[sigconf]{acmart}

\usepackage{soul}
\usepackage{xspace}
\usepackage{url}
\usepackage{graphicx}
\usepackage{listings,multicol}
\usepackage[caption=false]{subfig}  
\usepackage[export]{adjustbox}
\usepackage{amsmath}
\usepackage{textcomp}
\usepackage{breakurl}
\usepackage{listings}
\usepackage{cleveref}
\usepackage{algorithm}
\usepackage{enumitem}
\usepackage[noend]{algpseudocode}
\usepackage{multirow}

\newcommand{\todo}[1]{}

\newcommand{\ie}{\emph{i.e.,}\xspace}
\newcommand{\eg}{\emph{e.g.,}\xspace}

\newcommand{\toolname}{\texttt{Miner}\xspace}
\newcommand{\rminer}{\texttt{RefactoringMiner}\xspace}

\AtBeginDocument{%
  \providecommand\BibTeX{{%
    \normalfont B\kern-0.5em{\scshape i\kern-0.25em b}\kern-0.8em\TeX}}}

\begin{document}

\title{Assessing Project-Level Fine-Tuning of ML4SE Models}

\author{Egor Bogomolov}
\authornote{Both authors contributed equally to this research.}
\email{egor.bogomolov@jetbrains.com}
\affiliation{%
  \institution{JetBrains Research}
  \country{}
}

\author{Sergey Zhuravlev}
\authornotemark[1]
\email{sgzhuravlev@edu.hse.ru}
\affiliation{%
  \institution{Higher School of Economics}
  \country{}
}
\author{Egor Spirin}
\email{spirin.egor@gmail.com}
\affiliation{%
  \institution{JetBrains Research}
  \country{}
}
\author{Timofey Bryksin}
\email{timofey.bryksin@jetbrains.com}
\affiliation{%
  \institution{JetBrains Research}
  \country{}
}

\begin{abstract}

Machine Learning for Software Engineering (ML4SE) is an actively growing research area that focuses on methods that help programmers in their work. In order to apply the developed methods in practice, they need to achieve reasonable quality in order to help rather than distract developers. While the development of new approaches to code representation and data collection improves the overall quality of the models, it does not take into account the information that we can get from the project at hand.

In this work, we investigate how the model's quality can be improved if we target a specific project. We develop a framework to assess quality improvements that models can get after fine-tuning for the method name prediction task on a particular project. We evaluate three models of different complexity and compare their quality in three settings: trained on a large dataset of Java projects, further fine-tuned on the data from a particular project, and trained from scratch on this data. We show that per-project fine-tuning can greatly improve the models' quality as they capture the project's domain and naming conventions. We open-source the tool we used for data collection, as well as the code to run the experiments: \url{https://zenodo.org/record/6040745}.

\end{abstract}




\maketitle

\section{Introduction}\label{sec:introduction}

During recent years, research in applications of machine learning for software engineering (ML4SE) has grown in popularity. The goal of such applications is to ease the work of software developers by automating or assisting them in everyday tasks such as writing code~\cite{chen2021codex}, documentation~\cite{wang2021codet5}, finding bugs~\cite{hellendoorn2020great}, testing software~\cite{rongqi2021ml-tcp}, etc. Despite active work in the research community, the usage of ML4SE approaches in practice remains limited. One of the reasons for it is that most models do not show good enough quality for real-world usage. A possible solution to this problem is the adoption of very large pre-trained Transformer models like Codex~\cite{chen2021codex}. However, such models have billions (or even dozens of billions) parameters, and their real-time usage poses significant technical challenges. 

A common way to evaluate ML4SE approaches is to use a large dataset of open-source projects, extract data from them, and average the metrics describing the model's quality over all the data points collected from these projects. The number of projects in the testing dataset can be as large as hundreds, and if these projects are diverse enough, this kind of evaluation might give a pretty good understanding of the model's overall performance. However, when developers use a model to actually assist them in their work, they are less interested in the model's quality averaged over many projects. Instead, developers need the model's performance to be as high as possible on a specific project that they develop at the moment.

A possible way to improve the model's quality for a specific project is to integrate the knowledge about this project into the model. It can be done by fine-tuning the model on the data extracted from the project. Despite the straightforwardness of this idea, the benefits of per-project fine-tuning have not been investigated yet for ML4SE tasks. Existing works in the ML4SE domain either focus on project-specific models (\eg the ones that compute statistics inside the project~\cite{liu2018cc, Allamanis2015ngrams}), or large models that pre-train on large corpora of code~\cite{feng2020codebert,wang2021codet5}.

Correct assessment of models fine-tuned for a specific programming project turns out to be technically challenging. In a practical setting, model fine-tuning and inference are usually separated in time. Therefore, training and testing data for the per-project fine-tuning experiments should also be separated in time. However, for most tasks, existing datasets of source code include static snapshots of the projects (\ie state of the projects at a specific moment of time or at a specific commit)~\cite{Lu2021codexglue, husain2019codesearchnet}. Code samples from the project's static snapshot cannot be chronologically separated for training and testing due to the incremental nature of code changes in software projects. Without chronological separation, evaluation of fine-tuned models might give overly optimistic results as the models will be allowed to ``look into the future''. 
Thus, in order to correctly assess the quality of the per-project fine-tuning, we need to take the project's development history into account as well.

In this work, we introduce a novel data collection methodology that allows evaluation of machine learning models fine-tuned on the data from a specific project. Our methodology mimics the practical usage of such models: the data for fine-tuning comes from the project's snapshot, while validation and testing are done on method creations introduced chronologically after this snapshot. This way, we can assure that there is no data leakage between the samples used for fine-tuning and testing, \ie we ensure that during training the model did not access the data from the project's future. We implement the suggested approach as an open-source tool for mining datasets from the projects' histories. 


We then explore how the models' performance can be improved by targeting a specific project. For this purpose, we study the task of method name prediction, which requires a model to generate a method's name given its body and signature. According to the study by Minelli et al.~\cite{Minelli2015ide_usage}, developers spend 70\% of their time in integrated development environments (IDEs) reading code rather than writing it. For this reason, accurate and self-explanatory naming of methods and variables is very important~\cite{Butler2009naming}. The automated generation of better method names can assist developers in this task, but existing approaches to method name prediction do not yet achieve industry-level quality. Additionally, this task is popular in the ML4SE domain because it represents an extreme case of code summarization, which in turn can be seen as a benchmark of how good ML models are in code understanding~\cite{Allamanis2015ngrams, alon2018codeseq, Fernandes2019sns, zuegner2021codetransformer}. 

We compare three different models that work with different representation of source code, while all taking into account its underlying structure: CodeTransformer~\cite{zuegner2021codetransformer}, TreeLSTM~\cite{tai2015treelstm}, and Code2Seq~\cite{alon2018codeseq}. CodeTransformer is a Transformer-based~\cite{vaswani2017transformer} model that modifies computation of attention by adding information from the code's abstract syntax tree (AST). TreeLSTM is a modification of the LSTM~\cite{Hochreiter1997lstm} model that works with trees (ASTs in this case). Code2Seq is a model that works with path-based representation~\cite{alon2018pathbased} of code that proved to perform well for the code summarization task.
For evaluation, we use the \emph{Java-medium} dataset~\cite{alon2018codeseq}, which is commonly used in previous works on method name prediction. 

In particular, with CodeTransformer we analyze whether the information from a specific project improves the quality of popular large-scale pre-trained models. It turns out that CodeTransformer's F1 and ChrF scores can be improved by more than 15\% after fine-tuning on a particular project, while training from scratch on the data solely from the project quality-wise is comparable to pre-training on more than 15 million functions. TreeLSTM and Code2Seq also benefit from the per-project fine-tuning, even though it does not allow them to catch up with more complex and capable models.


With this work, we make the following contributions:
\begin{itemize}
    \item We suggest the idea of per-project fine-tuning of machine learning models working with source code. To the best of our knowledge, our work is the first to explore fine-tuning of ML4SE models in order to make them work better on particular projects.
    \item We develop a framework that makes possible the evaluation of models fine-tuned on specific projects, which we believe better represents the usage scenarios of the models in real life. The proposed approach takes the project's history into account and allows to separate the training, validation, and testing data in time.
    This way we ensure the absence of data leakages between training and testing data, \ie during training, the models do not have access to the information ``from the future'', which could happen in traditional evaluation approaches.
    \item We publish a tool that parses the development history of a software project written in Java and prepares a dataset for fine-tuning the models based on it and for their appropriate evaluation. In this study, we apply the tool to process projects from the existing \emph{Java-medium} dataset and assess the per-project fine-tuning of the models in the method name prediction task. However, the tool can be employed for data mining in other ML4SE tasks as well.
    \item We evaluate three models---CodeTransformer, Code2Seq, and TreeLSTM---in the method name prediction task. For each model, we assess three of its versions: without fine-tuning, with per-project fine-tuning, and trained from scratch on the project's data.
    We demonstrate that project-level fine-tuning might significantly boost the models' performance by enabling them to learn project-level conventions and domain-specific information. Importantly, it holds for the CodeTransformer model that was pre-trained on a very large data corpus. We demonstrate that by training only on the data from the project, for CodeTransformer, we can reach the quality similar to the pre-trained model without fine-tuning. 
\end{itemize}

Our findings highlight that while large-scale pre-training indeed boosts the models' quality, practical applications of ML4SE models might require taking available project-specific information into account.

\section{Assessing per-project fine-tuning}\label{sec:setup}

\begin{figure*}
  \centering
  \includegraphics[width=\textwidth]{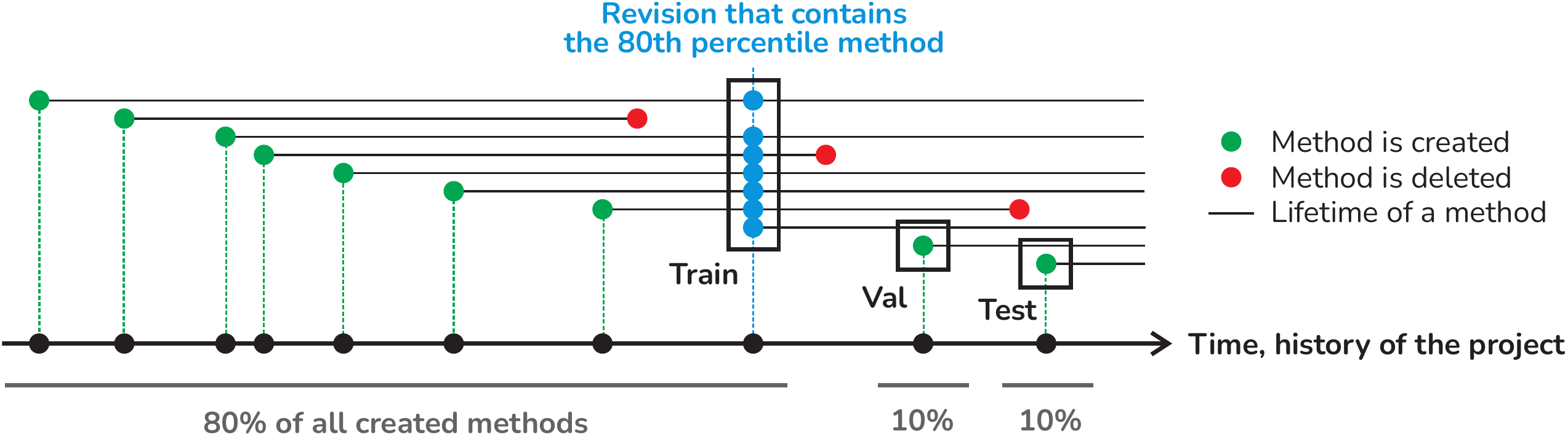}
  \caption{The pipeline for collecting the data. We collect all the information about the created methods in the history of a project, take the snapshot that contains the 80th percentile method, and use this snapshot's methods as the training set. The methods that were created after this snapshot are then divided evenly into the validation and testing sets.}
  \label{fig:pipeline}
\end{figure*}

\subsection{Motivation}

While the accuracy of predictions remains critical for practical applications of ML4SE models, existing works mostly focus either on models that gather information from a particular project where they will be applied~\cite{Allamanis2015ngrams, liu2018cc}, or training and evaluating models on separate large sets of projects~\cite{feng2020codebert, wang2021codet5}. We can combine both approaches by firstly training a model on a large code corpus and further fine-tuning it to capture project-specific information. This way, we can bring the models closer to the practical setting: in practice, programmers need the best possible model's quality for the project they develop at the moment rather than good average quality for an arbitrary set of projects. However, the correct assessment of the model's quality after fine-tuning could be challenging.

In this work, we study the task of method name prediction as it presents a popular benchmark of how well the models can understand the functionality of source code~\cite{zuegner2021codetransformer, alon2018codeseq, Fernandes2019sns}. A straightforward approach to test a fine-tuned model would be to randomly split the project methods into training, validation, and testing parts. However, this way we cannot assure that all the code fragments in the testing part were written after the ones from the training set. If this condition is violated, this could result in a data leakage that might lead to overly optimistic and incorrect results. For example, when programmers develop a method that uses some functionality that was not yet developed at the moment of fine-tuning, the model should not have any information about it. Hence, the model should be surprised by the input data in this case.
If we split all the project methods randomly, the model will likely capture usage examples of most methods and will be less surprised with the testing data than it would happen in practice.

While splitting training and testing data in time is a crucial element of the correct assessment of the model's quality after per-project fine-tuning, it might be also important for the general evaluation of ML4SE models. As time goes on, the code programmers write tends to change: programming languages add new features and code constructs, new libraries and frameworks emerge, popular technologies change. All of this leads to a drift in the input data for models working with source code. Nowadays, most works measure the quality of models by splitting the training and evaluation data \emph{at the project level}~\cite{zuegner2021codetransformer,husain2019codesearchnet}, hence the testing dataset might contain projects that were developed long before the ones in the training set. 
This evaluation technique---without taking the time aspect into account---might lead to the incorrectly high metric scores. We include the demonstration of such a case in the Appendix (see \Cref{sec:time-impact}). We expect this to be the case not only for the method name prediction task, but also for other software engineering problems, as the underlying reasoning is applicable to them as well.



\subsection{Proposed Evaluation Framework} \label{sec:dataset-mining}

In order to mitigate the aforementioned issues and correctly measure the effect of models' fine-tuning for a specific software project, we develop a novel approach that mimics the practical usage of a model that predicts method names. The developed framework can be extended to prepare data for other software engineering tasks that operate with code at method- or class-level: detection of outdated comments~\cite{liu2018automatic}, program repair~\cite{hellendoorn2020great}, recommendation of identifier names~\cite{Allamanis2015ngrams}, and others. The proposed approach ensures that the training part of the dataset consists of the methods that were implemented before the ones from the validation and testing sets. \Cref{fig:pipeline} presents the pipeline of data collection.

In the first step, we parse the Git history of a project and extract all \textit{method changes} from it. The difficult part of history processing comes from the fact that Git does not support proper handling of both file histories and methods' histories. Difficult cases arise when a method's signature was changed, a method was renamed, moved to another class, or the whole class containing the method was moved or renamed. Without explicit tracking, such situations would introduce two method changes: one that deletes the method, and one that creates it back, breaking the methods' histories.

To accurately track methods' histories, we propose to identify refactorings in each commit: renamings, relocations of methods and classes, method extractions, changes in method signatures, etc. We propose to use the information about such refactorings to extract method creations (\ie events when initial versions of methods are committed to Git), as we deal with the method name prediction task. However, other researchers can apply such an approach to mine data for other software engineering tasks: recommendation of names for identifiers~\cite{Allamanis2015ngrams}, detection of bug-introducing commits~\cite{hellendoorn2020great}, detection of outdated comments~\cite{liu2018automatic}, and others.


Once we extracted method creations from the project's history, we split them into two parts targeting the 80:20 ratio. Then we identify a commit that encompasses 80\% of the introduced methods (see \Cref{fig:pipeline} for a visualization), and use the project's snapshot at this commit to construct the training dataset. It resembles the way how programmers can use the data from the current version of the project to fine-tune a model and then use it for the subsequent project development. In the final step, we identify method creations that happened after the selected snapshot, and split them in two halves by the commits' date. The earlier half constitutes the validation part, while the later one constitutes the testing set. To further assure that there are no data leakages (\eg due to the presence of copied-and-pasted methods or code clones in the codebase), we deduplicate all the collected methods.

From the practical perspective, our framework mimics the way how the model can be fine-tuned on the data from a project at some point of its history and then be used to predict names or solve another SE task for the newly implemented methods as the project carries on. Since we split the methods by the time of their creation, we can be sure that there are no data leakages between the training and testing data. The developed approach can be beneficial not only for the method name prediction problem, but also for other ML4SE tasks targeting method granularity.

\subsection{\toolname}

We implement the described framework in a tool called \toolname. Currently, the tool supports working with Java, as it is the most commonly studied language in the ML4SE domain. \toolname is a plugin for IntelliJ IDEA written in Kotlin which runs the IDE in the headless mode (\ie without graphical user interface) and uses it to process Git history and extract data from it. In this, we follow the idea to use the IDE as a software analysis tool for mining data for ML4SE tasks which was initially suggested by Spirin et al.~\cite{spirin2021psiminer}.

In order to determine the refactorings performed in each change, we use \rminer~\cite{Tsantalis2020rminer2}, a library that identifies refactorings made in Java code in Git commits. The authors report 99.6\% precision and 94\% recall in the detection of more than 80 kinds of refactoring operations, which is the best performance among the refactoring detection tools. We employ the information about refactorings to identify renamings, moves, and extractions at method and class level. It further allows us to track the methods' history as the project develops further and to filter out all the method changes aside from their creations (\ie commits where the methods are initially written). 

After determining the project's snapshot (see \Cref{fig:pipeline}) that we use for per-project fine-tuning, we run deduplication to get rid of the methods in testing and validation sets that have duplicates in the training data. These duplicates can appear due to copying-and-pasting of functionality or they can be common simple methods such as getters or setters for variables with common names. We identify the duplicates with a two-step procedure. Firstly, we identify the exact textual duplicates: we remove all the comments and whitespaces from the methods and find the exact copies afterwards. Secondly, for each machine learning model that we study, we identify duplicates after extracting the model-specific code representation. Thus, if the methods have the same name and some subtle differences in their implementations that the model cannot see, we also remove them from the testing or validation data. This way, we ensure that the models' results are not overly optimistic.

\section{Experimental setup}\label{sec:ft-study-setup}

\subsection{Research Questions}\label{sec:rqs}

With this study, we aim to evaluate whether project-specific information can be helpful for machine learning models working with source code. In particular, we are interested in whether large-scale pre-training allows the models to mitigate the need of capturing project-specific information, and whether simpler models benefit more from the project's data. To investigate it, we formulate the following research questions:
\begin{itemize}
    
    \item \textbf{RQ1:} How does fine-tuning on the data from a specific project impact the quality of different machine learning models in the method name prediction task?
    
    \item \textbf{RQ2:} Can fine-tuning on the data from a specific project allow simpler models to catch up with the more complex ones?
    
    \item \textbf{RQ3:} Is information from the project alone enough to accurately predict method names?
    
    \item \textbf{RQ4:} What do models learn from the project's data?
    
\end{itemize}


With RQ1, we aim to study whether fine-tuning on the project's data can improve results of the models with different complexity.

By answering RQ2, we want to understand whether fine-tuning can help simpler models to overcome their more complex counterparts. If that is the case, the development of models that are better suited for handling project-specific data (even by trading off other capabilities) might be of interest. 

Answering RQ3, we study models that did not see any data other than from the project itself. Training of such models is way faster compared to the pre-training on vast amounts of data, which is important in a practical setting. 

RQ4 targets the details models learn after fine-tuning on a project (\eg project's naming conventions, domain of the project, usages of particular libraries, etc.). This information can pave the way to enhancing the models by explicitly including the needed data into the model's input, which can be seen as an alternative to fine-tuning. 

\subsection{Evaluation Scheme}\label{sec:evaluation-scheme}

\textbf{Method name prediction task.} In prior works, researchers formulate the task of method name prediction as follows: given the whole method implementation (including its return type, modifiers, arguments, and body) with its name being replaced with a stub token, the model should generate the original method's name. The method itself in this case can be represented as a sequence of text tokens~\cite{feng2020codebert}, abstract syntax tree~\cite{alon2018codeseq}, a complex graph~\cite{Fernandes2019sns}, etc.
While some earlier works on this problem formulated it as a classification task (\ie every unique method name belongs to a separate class)~\cite{alon2019code2vec}, this approach does not generalize well. Consecutive studies treated the problem as a task of sequence generation~\cite{zuegner2021codetransformer, Fernandes2019sns}. A method name in this case is split into a sequence of sub-tokens by CamelCase or snake\_case. This approach allows models to generate names that they did not encounter in the training set.

\textbf{Metrics.} Most studies that evaluated models for method name prediction used F1-score to compare the results~\cite{Allamanis2015ngrams}. Its computation relies on the fact that models generate method names as sequences of sub-tokens. Precision is then computed as the percentage of sub-tokens in the generated sequence that also appear in the actual method's name. Recall is the percentage of sub-tokens in the original method's name that appear in the prediction. F1-score is the harmonic mean of precision and recall. Thus, this metric ignores the order of sub-tokens in the prediction. It is motivated by the method names like \texttt{linesCount} and \texttt{countLines}, where changes in the word order do not affect the meaning.

A recent work by Roy et al.~\cite{roy2021} studied the applicability of various metrics from the natural language processing (NLP) domain for code summarization tasks. The authors evaluated code summarization models with BLEU~\cite{papineni2002bleu}, METEOR~\cite{banerjee2005meteor}, ROUGE~\cite{lin2004rouge}, ChrF~\cite{popovic2015chrf}, and BERTScore~\cite{zhang2020bertscore}, and compared the results with human assessment. They conclude that METEOR and ChrF are the most appropriate metrics for the code summarization task. As the method name prediction is often thought of as an ``extreme summarization of code'' (\ie the model needs to describe the method's functionality in a very concise way), we also choose ChrF as another metric that we look at in this study.

\textbf{Cleaning sub-token sequences.} Despite the splitting of method names into sub-tokens, some sub-tokens in the testing part of the dataset might be out of the model's vocabulary. In this case, we do not remove them from the sequence when computing F1 and ChrF scores as it points at the model's incapability of generating proper method names in these particular cases. 
The other way around, when models generate a special <UNK> sub-token that corresponds to out-of-vocabulary words, we remove it from the generated sequence as in practice it will not be displayed to a user. 

Finally, we remove repeated sub-tokens from the computation of the metric scores, since otherwise they might skew the metrics. To understand the motivation for such filtering, consider the following example: the original method name is \texttt{getApple} and a model generates \texttt{getGetGetGetOrange}. If we do not remove the repeating sub-tokens, we might consider the model's precision to be 80\% as it generated the word \texttt{get} from the correct sequence multiple times. However, the users will most likely treat such prediction to be way worse than a simple \texttt{getOrange} that scores 50\% precision.

\textbf{Model comparison.} In order to compare different models, we run paired bootstrap tests~\cite{efron1994introduction} over their predictions for each test project. A detailed description of this test can be found in the work by Dror et al.~\cite{dror2018guide}. In short, given the results of two models for the samples from a single project, we resample the result pairs with replacement 10 thousand times, and for each resampled set, we determine which model has the better average metric on this set. The probability of one model beating the other for a specific project is then the ratio of resamples where it has a better score. We consider one model to confidently beat another on a project when the probability of it being the winner is higher than 95\%.

\subsection{Dataset}\label{sec:dataset}

For evaluation of the models that were fine-tuned per-project, we use the \emph{Java-medium} dataset. It was introduced by Alon et al. and was then used to benchmark method name prediction models~\cite{alon2018codeseq}. The dataset includes 1,000 top-starred Java projects from GitHub. The training part of the dataset includes 800 projects with 100 projects per validation and testing sets each. When the dataset was originally introduced, it contained about 3.8M methods. We collected the up-to-date versions of these projects from GitHub in August 2021. Four projects from the original list turned out to be deleted or made private. 

For the method name prediction task, the methods should be filtered to avoid meaningless examples and data leakages. Following the previous works~\cite{alon2018codeseq,spirin2021psiminer}, we apply the following filtering to ensure data quality:
\begin{itemize}
    \item Mask method names in recursive calls. Otherwise, the model will learn to identify recursive calls and copy method names from them.
    \item Remove empty methods. For such methods, models can only random guess the answer, which skews the results. Moreover, generating a name for an empty method looks like a rather impractical scenario.
    \item Remove abstract and overridden methods. Such methods contribute multiple samples with the same label to the dataset, virtually giving the corresponding names higher weight when training the model. However, at the time of applying the model in practice, it is unlikely that an abstract method can be correctly named based on a single overridden instance. Thus, we decided to remove both abstract and overridden methods.
    \item Remove constructors. Their names are identical to their class names, but generation of a class name based on a single constructor seems impractical.
    \item Remove comments and javadocs. They might contain hints at the current name of the method, while the goal of the task is to summarize the method's functionality.
\end{itemize}

We use PSIMiner~\cite{spirin2021psiminer} to filter methods and extract training data for the models (see \Cref{sec:models}): tokenized code, abstract syntax trees, and path-based representation of code.

We ran \toolname for all the projects in the testing set of \emph{Java-medium}. We removed the projects where the number of samples in the testing dataset after filtering was less than 20. Such a small number of samples would not allow a statistically significant model comparison. We manually explored the reasons why such projects appear in the dataset and found three main reasons for that:
\begin{itemize}
    \item a project itself is just too small;
    \item we filter out almost all the methods: it happened for some of the Android projects that only contained overridden methods from the built-in or library classes;
    \item a project contains incomplete Git history: while the project is large, almost all the code was pushed to the repository in the initial commit.
\end{itemize}

In the end, out of the 100 projects of the testing part of the \emph{Java-medium} dataset, 47 of them made it to our resulting dataset, containing at least 20 testing samples each. We include a full list of the studied projects in the Appendix (see \Cref{tab:list-projects}). 

\subsection{Models}\label{sec:models}

In this work, we compare three machine learning models that were previously applied to the method name prediction task, namely CodeTransformer, Code2Seq, and TreeLSTM. With this, we aimed to select a set of models that is diverse from the point of their architecture and complexity. We describe the models in more detail in the Appendix (see \Cref{app:models}). In this section, we provide their brief overview.

\textbf{CodeTransformer.} CodeTransformer~\cite{zhang2020bertscore}, a large Transformer-based model that is adapted to working with code by taking into account its abstract syntax tree. In contrast to other Transformer-based models working with code (\eg CodeT5~\cite{wang2021codet5}, CodeBERT~\cite{feng2020codebert}), CodeTransformer was previously evaluated on the method name prediction task. In this work, we use the pre-trained model published by the authors. The model contains about 108M parameters.

\textbf{Code2Seq.} Code2seq~\cite{alon2018codeseq} is a code-specific model introduced by Alon et al. that works with the path-based representation of code~\cite{alon2018pathbased}. By including Code2Seq in the study, we aim to answer RQ2 --- whether additional fine-tuning will allow shortening the gap between this model and CodeTransformer that benefited from large pre-training. We use our own implementation of the model. We train the model on the \emph{Java-medium} dataset achieving the same quality on its testing part as was reported by Alon et al. The total size of the model is 7.5M parameters.

\textbf{TreeLSTM.} TreeLSTM~\cite{tai2015treelstm} is a simpler model compared to the previous two. It was originally designed to work with trees built from sentences in natural languages. However, it was also applied to source code authorship attribution~\cite{alsulami2017}, method name prediction~\cite{alon2018codeseq}, and code generation~\cite{Liu2018TreeGANSS}. With TreeLSTM, we study whether a simpler model can improve the quality by fine-tuning to catch up with the modern architectures. We train this model on the \emph{Java-medium} dataset for the method name prediction task. On the testing part of \emph{Java-medium}, we achieve the quality similar to the one reported in other papers. The total size of our TreeLSTM is 6.2M parameters.

\subsection{Project-level Training}


The models described previously are trained to maximize the quality across all the methods in the validation set, no matter what project they come from. In this work, we refer to such training setup as ``Original''. We evaluate the pre-trained models on the testing data collected from each project (see \Cref{sec:dataset}).

In order to analyze the impact of per-project fine-tuning, we fine-tune the pre-trained version of each model on the data from each project (\ie for 47 projects, we get 47 fine-tuned versions of each model). We refer to the resulting models as ``Fine-tuned''.

Finally, to understand whether the information solely from a particular project is enough to make decent predictions, we train each model from scratch on the data from each project (\ie for 47 projects, we get 47 versions of each model trained from scratch). We refer to this setup as ``Project-only''.

\section{Results}\label{sec:experimental-results}

\begin{figure*}
  \centering
  \includegraphics[width=\textwidth]{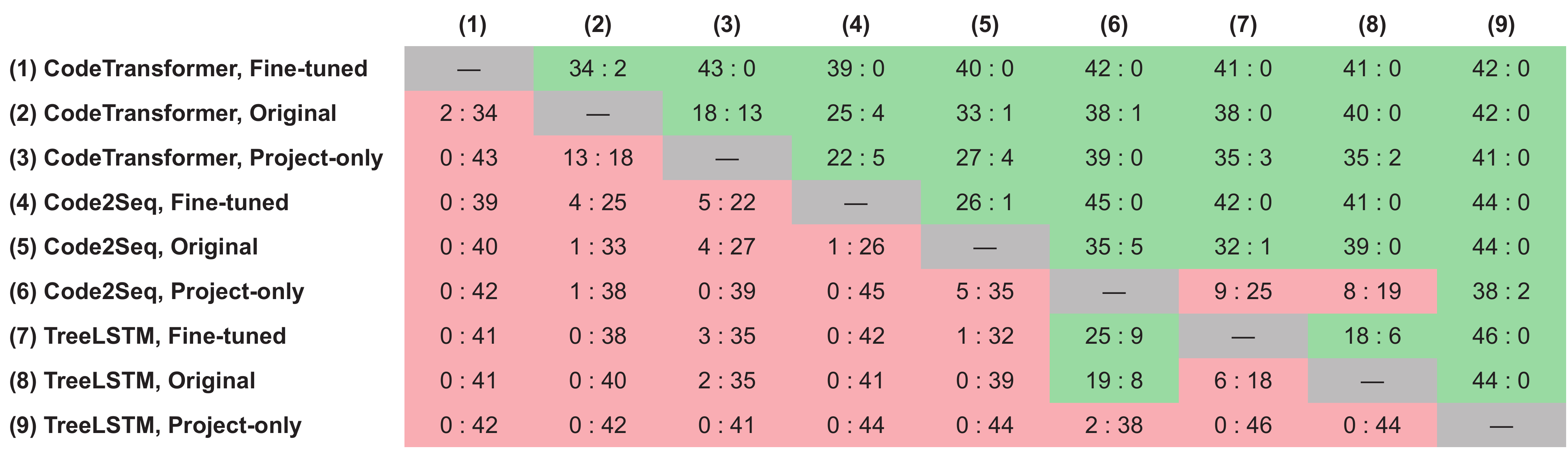}
  \caption{Results of paired bootstrap tests for all pairs of models on projects with at least 20 testing samples (47 projects in total). Numbers $N_A:N_B$ in row $A$ and column $B$ mean that model $A$ performs significantly better than $B$ for $N_A$ projects and significantly worse for $N_B$ projects. For the remaining projects statistical test cannot confidently distinguish the models. For example, CodeTransformer after fine-tuning performs significantly better than its pre-trained version for 34 projects, while for 2 it performs worse. For the remaining 11 projects (47-34-2) the difference is insignificant.}
  \label{fig:bootstraps}
\end{figure*}

\begin{table}[b]

\caption{Averaged F1 and ChrF scores for all the studied models. Large projects contain at least 100 testing samples, small ones --- between 20 and 100 testing samples.  ``ORIG (Original)'' models trained to solve the method name prediction task without seeing the data from the test project. ``FT (Fine-tuned)'' models were further trained on the data from a specific project. ``PO (Project-only)'' models trained only on methods from one project.}
\label{tab:all-results}
\centering
\begin{tabular}{cccccc}
\toprule
\multicolumn{2}{c}{\multirow{2}{*}{\textbf{Model}}}                                               & \multicolumn{2}{c}{\textbf{Large projects}} & \multicolumn{2}{c}{\textbf{Small projects}} \\\cmidrule(lr){3-4}\cmidrule(lr){5-6}
                                 &              & \textbf{F1}               & \textbf{ChrF}             & \textbf{F1}               & \textbf{ChrF}            \\ \midrule
\multirow{3}{*}{CodeTransformer} & FT   & \textbf{47.6}             & \textbf{45.0}             & \textbf{57.7}             & \textbf{55.1}            \\
                                 & ORIG     & 40.9             & 39.4             & 38.6             & 37.7            \\
                                 & PO & 42.9             & 40.4             & 43.1             & 39.3            \\ \midrule
\multirow{3}{*}{Code2seq}        & FT   & \textbf{40.2}             & \textbf{32.6}             & \textbf{44.1}             & \textbf{37.4}            \\
                                 & ORIG     & 34.5             & 28.0             & 37.4             & 31.4            \\
                                 & PO & 30.8             & 23.9             & 24.2             & 19.5            \\ \midrule
\multirow{3}{*}{TreeLSTM}        & FT   & \textbf{29.2}             & \textbf{23.1}             & \textbf{34.2}             & \textbf{29.2}            \\
                                 & ORIG     & 26.6             & 21.2             & 31.6             & 26.2            \\
                                 & PO & 15.4             & 12.0             & 12.5             & 10.7        \\  \bottomrule 
\end{tabular}
\end{table}

In this work, we use the proposed evaluation framework for per-project fine-tuning to compare three ML4SE models: CodeTransformer, Code2Seq, and TreeLSTM.
For each model, we apply three different scenarios to train them: ``Original'', ``Fine-tuned'', and ``Project-only''.
Thus, we have nine different model variants for comparison. In order to analyze the experimental results, we report average ChrF and F1 for each model variant (see~\Cref{tab:all-results}) and pairwise statistical comparison with other model variants (see~\Cref{fig:bootstraps}). We also report an average ChrF for each project with at least 100 testing samples in the Appendix (see \Cref{fig:large_projects}).

\smallskip
\textbf{RQ1: How does fine-tuning on the data from a specific project impact the quality of different machine learning models~in the method name prediction task?}

In order to answer the first research question, we perform a paired bootstrap test to compare the results of the studied models with and without fine-tuning (see \Cref{fig:bootstraps}). For this, we consider projects with at least 20 testing samples.

For CodeTransformer, fine-tuning leads to a statistically significant increase in the model's performance for 34 projects, while for 11 we observe no difference. For two projects, fine-tuning actually worsens the results. It happens because in our evaluation framework, training, validation, and testing data are separated in time. Therefore, for some projects there is a discrepancy between the validation and testing data (\eg due to changes in the project development) which leads to sub-optimal early stopping.
Judging by these results, we conclude that fine-tuning leads to better CodeTransformer's quality for the majority of the projects. On average, the F1-score of the model changes from 40.9 to 47.6 for large projects (see \Cref{tab:all-results}), which is a relative increase of more than 15\%. Similarly, ChrF improves from 39.4 to 45.0. 

For Code2Seq, fine-tuning significantly improves the results for 26 projects, worsens for a single one, and does not change the quality for 20 projects. The average F1-score on large projects increases from 34.5 to 40.2, which is a relative growth similar to what we see for CodeTransformer. Overall, fine-tuning allows Code2Seq to capture helpful information for the majority of the projects.

Finally, for TreeLSTM, we get significant improvements for 18 projects but a decrease in results for 6 of them. For 23 projects, there are no significant changes. On average, the F1-score changes from 26.6 to 29.2, which is a less than 10\% relative increase. Both the per-project results and improvements in the metric scores turn out to be smaller for TreeLSTM compared to other models.

Our experiments show that the effect of fine-tuning does not decrease as models become larger and more complex. In contrast to other models that we study, CodeTransformer has an order of magnitude more parameters (more than 100 million) and is pre-trained on a much larger dataset. It was first trained to solve the permutation language modeling task and then to predict method names on more than 15 million Java methods. Despite the massive pre-training, the model still can derive useful information from the project's data. Fine-tuning on the project-specific data allows to (1) significantly improve the model's performance for most of the projects and (2) get an even higher relative increase in F1-score than for simpler models. Thus, we conclude that per-project fine-tuning can be a useful practice even for modern large-scale models. 

\smallskip
\textbf{RQ2: Can fine-tuning on the data from a specific project allow simpler models to catch up with the more complex ones?}

For the second research question, we compare the performance of the fine-tuned Code2Seq and TreeLSTM models to the pre-trained version of CodeTransformer, as well as the fine-tuned TreeLSTM to the pre-trained Code2Seq. \Cref{fig:bootstraps} shows the results of the models' pairwise comparison for all the projects.

As we described in the first research question, TreeLSTM cannot learn much new information from the project-specific data. Given that, it cannot catch up neither with Code2Seq nor CodeTransformer, as even after fine-tuning it performs significantly worse for most of the projects. The only project where it can outperform Code2Seq is \emph{talon-twitter-halon}. It happens because the project contains a lot of methods that have the same names in both testing and training data. Such methods in this project have slight differences in implementation, which TreeLSTM can successfully ignore.

Even though Code2Seq can beat the pre-trained version of CodeTransformer on 4 projects and performs of par on 18 of them, it still shows significantly worse quality on 25 projects. Moreover, for the projects where Code2Seq manages to catch up, the fine-tuned version of CodeTransformer consistently achieves better average F1-score and ChrF, most of the time with statistical significance.

Based on the investigation in this and previous research questions, we conclude that fine-tuning is more beneficial for more complex and better-trained models. Even though simpler models also benefit from fine-tuning, it does not allow them to close the gap in quality with the more complex ones.
    
\smallskip
\textbf{RQ3: Is information from the project alone enough to accurately predict method names?}

We perform a set of experiments where the models are trained from scratch on the data from a specific project. We initialized the models' weights with random values the same way as it was done during pre-training. Similar to the fine-tuning experiments, CodeTransformer and Code2seq benefit from the project's data, while TreeLSTM cannot generalize well based on it. \Cref{fig:bootstraps} shows results of the comparison for the models trained from scratch.

According to the paired tests, for TreeLSTM, the model trained on the data solely from one project performs worse than the pre-trained version for 44 projects and achieves matching results only for 3. For Code2Seq, the results are similar: most of the time, the project-only version is significantly worse than the pre-trained one.

Surprisingly, the project-only version of CodeTransformer performs comparable to the pre-trained version of the model. The project-only model has better quality for 13 projects, worse quality for 18, and the models are indistinguishable 16 times. For some projects, project-only training achieves substantially higher results: for the \emph{rest-assured} project the average ChrF nearly doubled. Also, the project-only CodeTransformer consistently outperforms Code2Seq.

Good performance of CodeTransformer trained on the project's data is an interesting result since model training on samples from a single project is quicker than its pre-training on thousands of projects. It suggests that the study of complex models trained on just a single project can be an interesting research direction as it allows cutting computation time.
However, a combination of pre-training with further fine-tuning remains a better option: the project-only CodeTransformer can achieve statistically similar performance only for 4 projects and falls behind for 43 of them.

\smallskip
\textbf{RQ4: What do models learn from the project's data?}

We examined predictions of the models before and after fine-tuning in order to identify patterns in the cases where fine-tuning improved the predictions. One source of the improvement is the presence of methods with the same names in the testing and fine-tuning data mined from a project. We should note that methods with the same names are never completely identical, as we carefully track methods' histories and also run additional deduplication to account for copied-and-pasted methods. Based on our analysis, such cases arise because of the three major reasons:

    \textit{Common methods:} for example, getters and setters with more complex logic than just returning a variable. Complex logic leads to implementation differences between methods, \eg calls to different types of data providers. 
    
    \textit{Architectural flaws:} in some projects, we encountered many classes with similar roles and functionality (\eg data providers, helpers, etc.) that do not inherit from a common abstract class or implement a common interface. It leads to the creation of methods with the same names and similar logic in each class, which would otherwise be marked as overridden. In particular, such a situation happened in the \emph{talon-twitter-holo} project that shows the biggest boost in performance after fine-tuning.
    
    \textit{Tests:} large projects have naming conventions for tests. Test methods for classes with similar logic (\eg implementing a common interface) test similar functionality, so they often share names.

Methods with names that appeared in the fine-tuning dataset constitute about 33\% of samples where the fine-tuned CodeTransformer model performs better than the pre-trained one. 
For the other two models, the respective numbers are 28\% for Code2Seq and 25\% for TreeLSTM. Thus, models improve not only because they saw similar methods in the fine-tuning data. Moreover, the identified cases are important for models to solve as they keep the names consistent across the project.

Another reason for improvement after fine-tuning is learning of project-level naming conventions. 
For example, the names of most test methods in the \emph{gocd} project start with the word ``should'' rather than with the commonly used ``test''. The fine-tuned models successfully capture this information: for CodeTransformer, on \emph{gocd}, roughly 60\% of the generated names that improved in quality after fine-tuning are tests starting with ``should''. The pre-trained model identifies test methods but suggests names starting with ``test''. A similar situation happens for the \emph{rest-assured} project, where test methods start with ``can''. Thus, fine-tuned models are better at following project-level conventions.

Also, fine-tuning allows the models to focus on the project's domain. By training on a huge dataset, models maximize average quality. For the \emph{Java-medium} dataset, the most popular domain is Android. Thus, the models seem to follow the Android naming conventions. For example, in the \emph{MinecraftForge} project, the pre-trained version of CodeTransformer generates the name ``subscribe'' for some methods that manipulate rendering logic. These methods have the words ``event'' and ``register'' close to each other, which confuses the model. We observe that per-project fine-tuning allows the models to move from general names to more subtle domain-specific logic.

\section{Discussion and Future Work}

The experimental results suggest that per-project fine-tuning can improve the quality of various models. Although CodeTransformer is a model pre-trained on a huge dataset of Java code, its F1-score can be improved by more than 15\% by capturing information from the project at hand. Fine-tuning does not have more impact for simpler models than for the complex ones, so it cannot bridge the gap between them. Thus, in order to bring the ML-based approaches to method name prediction closer to practical usage, a promising direction is training of large models and their further per-project fine-tuning.

Interestingly, for CodeTransformer we achieve similar results for both the pre-trained model and the one trained from scratch on the data from a specific projects. Since the number of samples in a single project, even a large one, is orders of magnitude smaller than in the pre-training dataset, training of the models solely on a project might be a topic that needs a deeper further investigation.

We identify three major patterns that models tend to capture from the per-project fine-tuning: methods with repetitive names, project-level naming conventions, and project's domain. An important feature required for practical applications of the method name prediction models is the ability to keep the names consistent within the project. The correct usage of repetitive names and project-level naming conventions is a crucial step for naming consistency. As fine-tuning on the project's data allows capturing such information, it makes models more practically applicable. Searching for other ways of integrating the information required for keeping naming consistency becomes an interesting topic for future research.

\section{Related work}\label{sec:related-work}

Aside from the models that we study in our work, researchers used other approaches to solve the method name prediction task. Allamanis et al.~\cite{Allamanis2015ngrams} suggested an approach based on N-grams that works at the project level. Alon et al. suggested Code2Vec~\cite{alon2019code2vec}, a simpler ML model that builds on top of path-based representation of code. 
Fernandes et al.~\cite{Fernandes2019sns} used gated graph neural network (GGNN)~\cite{ggnn} to predict method names. GGNN allows working with more complex graph representation of code than with AST: the authors combined AST with data flow graph, control flow graph, and added augmenting edges. While the work showed results superior to the previously reported ones, it only evaluated on the Java-small dataset consisting of 11 projects. In terms of the raw quality, the suggested model was later surpassed by CodeTransformer. Evaluation of GGNN and other graph-based models working with source code~\cite{hellendoorn2020great} in the per-project fine-tuning setting is an important topic for the future work.

Recently, researchers suggested many different Transformer-based models for working with software code: CodeTransformer~\cite{zuegner2021codetransformer}, CodeT5~\cite{wang2021codet5}, CodeTrans~\cite{elnaggar2021codetrans}, CodeBERT~\cite{feng2020codebert}, Codex~\cite{chen2021codex}. Out of these models, only CodeTransformer was evaluated for the method name prediction task. Therefore, we evaluate it in our work, as application of other models requires their adaptation to the new task and further re-training.

\section{Conclusion}\label{sec:conclusion}

Machine learning on source code is a promising research direction that gains a lot of attention recently with the works like Codex~\cite{chen2021codex} and AlphaCode~\cite{li2022alphacode}. In this work, we suggest to fine-tune the ML models working with source code on the data from specific software projects. While the common practice is to train models on large datasets of code, in practice programmers need the best possible model's quality for a project that they are working with at the moment. By fine-tuning on the data from this project, we can bring ML models closer to practical applicability.

In our work, we target the method name prediction task that aims to generate high-quality names for code methods. While the task serves as a popular benchmark for assessing models' ability to understand code functionality, it also has practical value for keeping the method names consistent and self-explainable. 
We develop an evaluation framework that allows correct assessment of model's quality before and after fine-tuning on a specific project. It takes into account the project's history of development in order to separate training, validation, and testing data in time. We also implement the developed framework as an open-source tool for automated mining of datasets from arbitrary Git projects. We applied the tool to mine the data from the commonly used \emph{Java-medium} dataset, leaving us with 47 preprocessed projects for further tests.

We then use the developed framework and the collected data to evaluate three models: CodeTransformer, Code2Seq, and TreeLSTM. We compare the results of their pre-trained and fine-tuned versions, as well as train them from scratch on the data solely from a single project. Our experiments for 47 projects suggest that all the models benefit from fine-tuning, but more complex ones can capture even more information this way, despite the larger pre-training and capacity. For CodeTransformer, we achieve a 15\% relative improvement in F1 and ChrF scores with statistically significant improvement of the results on almost all the test projects.

In order to ease the research in the direction of per-project fine-tuning, we open-source \toolname---the tool for mining evaluation data for per-project fine-tuning---as well as the code for running our experiments: \url{https://zenodo.org/record/6040745}. All the collected datasets will be published upon paper acceptance.

\bibliography{paper}
\bibliographystyle{ACM-Reference-Format}

\appendix

\section{Models}\label{app:models}

In this section, we describe the models used for comparison in more detail.

\subsection{CodeTransformer}

CodeTransformer~\cite{zhang2020bertscore} is a large Transformer-based model that is adapted to working with code by taking into account its abstract syntax tree. While recently there were other large Transformer-based models trained and applied on source code (\eg CodeT5~\cite{wang2021codet5}, CodeBERT~\cite{feng2020codebert}), CodeTransformer is among the few that both (1) take code structure into account and (2) were evaluated on the method name prediction task.

The main distinguishing feature of this approach is incorporating relative distances between nodes in the AST.
In the original Transformer model~\cite{vaswani2017transformer}, authors proposed to embed information about the position of each token into the model.
Later, Dai et al.~\cite{dai2019transformerxl} proposed to replace absolute positions with relative positions between tokens in the input sequence. The authors of CodeTransformer transfer this idea to source code, but as relative distances between tokens, they use distances between the corresponding nodes in the code's AST. CodeTransformer simultaneously uses four different types of distances between the nodes in the tree: shortest path length, ancestor distance, sibling distance, and Personalized PageRank~\cite{page1999ppr}.
As a part of the training process, the authors apply large pre-training with the permutation language modeling objective~\cite{yang2019xlnet} on a combination of the \emph{Java-medium} and Java-large datasets~\cite{alon2018codeseq}, containing more than 15 million methods in total. The pre-trained model is further fine-tuned for the method name prediction task on the Java-small dataset. Currently, CodeTransformer shows one of the top results in method name prediction.



\subsection{Code2Seq}

Code2seq~\cite{alon2018codeseq} is a code-specific model introduced by Alon et al. that works with the path-based representation of code~\cite{alon2018pathbased}. The model was originally applied for the method name prediction task and since then became a strong baseline for other works targeting this task. Among the models that do not use large-scale pretraining on unlabeled data, Code2Seq remains one of the best models that use no other data except for the code's AST.

The path-based representation of code used by Code2Seq is built in the following way. 
For each snippet of code, a set of path contexts is extracted from the AST.
Each path context is a triple of tokens in two AST leaves and a sequence of node types along the path between them.
From the model architecture's perspective, Code2Seq at first transforms each path context into a numerical vector by folding the sequence of node types with LSTM~\cite{Hochreiter1997lstm} and adding embeddings of tokens.
After that, to generate a target sequence, Code2Seq uses LSTM with Luong's attention~\cite{luong2015attention} over the path-contexts' embeddings. 

In the original work~\cite{alon2018codeseq}, the authors applied Code2Seq to the method name prediction and documentation generation tasks.
They used the described approach to train the model on Java-small, \emph{Java-medium}, and Java-large datasets and reported state-of-the-art results at that time.
Nowadays, this model is still very popular as a strong baseline due to a good combination of quality, simplicity, and speed.



\subsection{TreeLSTM}

TreeLSTM~\cite{tai2015treelstm} is a simpler model compared to the previous two. It was originally designed to work with trees built from sentences in natural languages. However, it was also applied to source code authorship attribution~\cite{alsulami2017}, method name prediction~\cite{alon2018codeseq}, and code generation~\cite{Liu2018TreeGANSS}. With TreeLSTM, we study whether a simpler model can improve the quality by fine-tuning to catch up with modern architectures.

The model proposed by Tai et al. is based on a regular LSTM model~\cite{Hochreiter1997lstm}, modified to work with tree-like input structures.
The original LSTM cell works with sequences --- it recurrently processes each token with the proposed gates mechanism.
When applying the model to a tree, the authors suggest traversing it in the topological order from leaves to the root and applying a modified LSTM cell to each node. The node uses information from its children as input and passes the information to its parent.
Since this way cells need to work with more than one input, the authors proposed two possible modifications: an $N$-ary cell for trees with the constant number of children and \emph{ChildSum} for the rest. The latter is a cell that sums up its children's representations into a single vector and processes it with an LSTM cell without any further changes. For the AST, only the ChildSum cell is applicable.


\begin{table}[t]
    \caption{The list of 47 GitHub projects that we use for evaluation. The projects contain at least 20 testing samples each after data extraction.}
    \label{tab:list-projects}
    \centering
    \begin{tabular}{cc}
\texttt{Activiti} & \texttt{graphql-java} \\
\texttt{AmazeFileManager} & \texttt{hibernate-orm} \\
\texttt{CameraFragment} & \texttt{hover} \\
\texttt{DiscreteScrollView} & \texttt{http-kit} \\
\texttt{ExoPlayer} & \texttt{incubator-heron} \\
\texttt{Fragmentation} & \texttt{interview} \\
\texttt{GestureViews} & \texttt{jmeter} \\
\texttt{MSEC} & \texttt{keycloak} \\
\texttt{MinecraftForge} & \texttt{keywhiz} \\
\texttt{NoHttp} & \texttt{languagetool} \\
\texttt{Smack} & \texttt{lottie-android} \\
\texttt{Small} & \texttt{objectbox-java} \\
\texttt{StickyHeaderListView} & \texttt{requery} \\
\texttt{VasSonic} & \texttt{rest-assured} \\
\texttt{VirtualAPK} & \texttt{robotium} \\
\texttt{android-advancedrecyclerview} & \texttt{socket.io-client-java} \\
\texttt{android-priority-jobqueue} & \texttt{solo} \\
\texttt{btrace} & \texttt{spring-boot-starter} \\
\texttt{dropwizard} & \texttt{talon-twitter-holo} \\
\texttt{eureka} & \texttt{wildfly} \\
\texttt{feign} & \texttt{xUtils3} \\
\texttt{flow} & \texttt{xabber-android} \\
\texttt{generator} & \texttt{zipkin} \\
\texttt{gocd} & \texttt{} \\
    \end{tabular}
\end{table}

\section{Detailed Results}

\begin{figure}[tbp]
    \centering
    
    
    \includegraphics[width=\columnwidth]{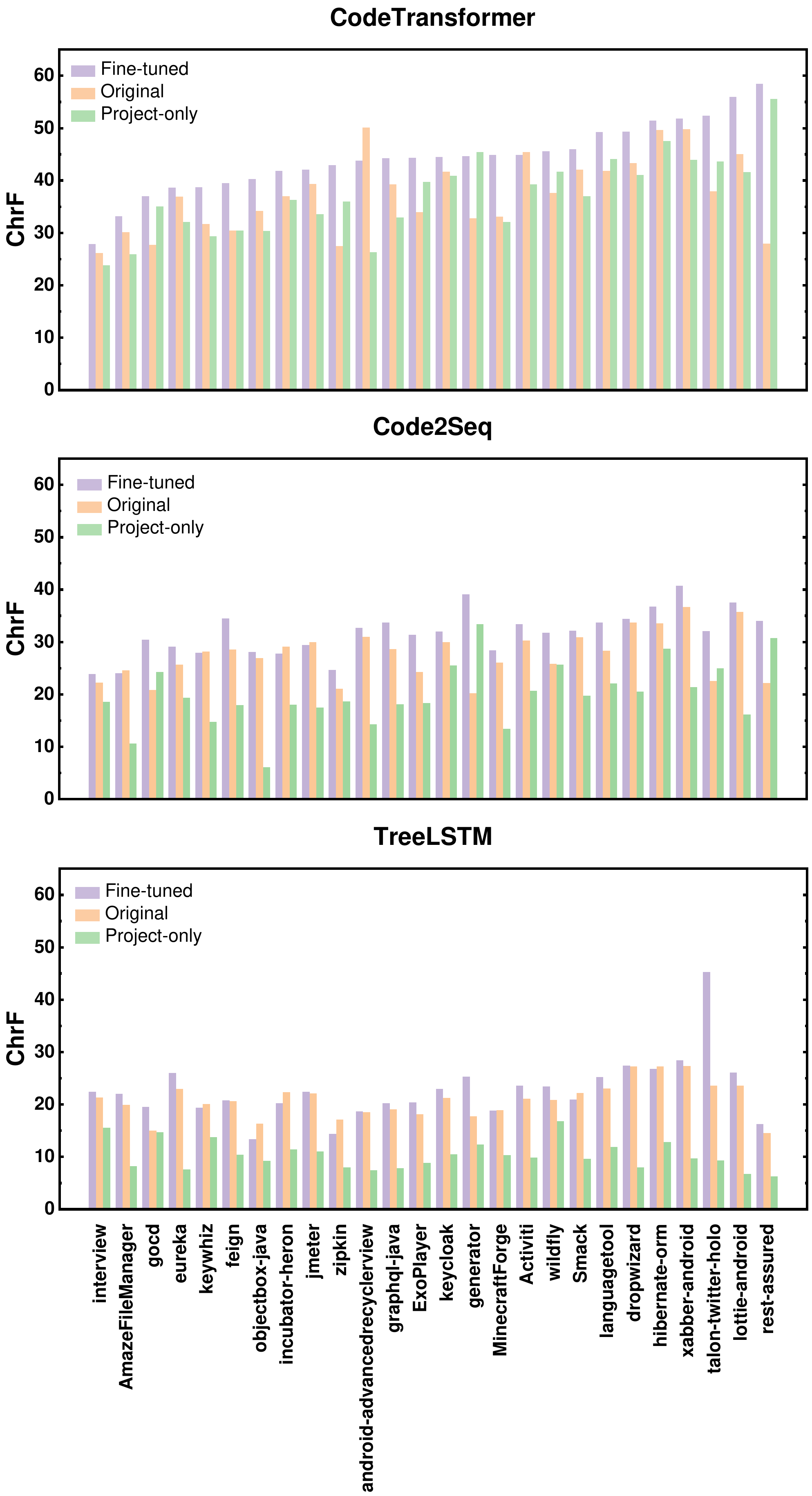}
    
    \caption{Average ChrF metric for predictions of CodeTransformer, Code2Seq, and TreeLSTM on projects with at least 100 testing samples. ``Original'' is a pre-trained model, ``Fine-tuned'' is a pre-trained model fine-tuned on project-specific data, ``Project only'' is a model trained from scratch solely on data from a project.}
    \label{fig:large_projects}
\end{figure}

Here we present a more detailed view of the results we get for projects with at least 100 testing samples. In \Cref{fig:large_projects}, you can see the results of the studied models on each of 26 large projects. We report average ChrF values for three versions of each model: pre-trained, fine-tuned on the project's data, and trained from scratch on data from the project. \Cref{tab:list-projects} shows the list of the 47 projects that we use to evaluate models. The projects come from the testing part of \emph{Java-medium} dataset. Each project contains at least 20 testing samples after applying the data extraction procedure described in \Cref{sec:dataset-mining}.

\section{Time Impact on the Evaluation}\label{sec:time-impact}

In order to ensure that the implementations of the models we use in our work are correct, we evaluated them on the testing part of the \emph{Java-medium} dataset. The results we got for all the models match with the results reported in the respective papers up to the small differences in data filtering and metric computation. However, when testing the models on the newly collected data (method creations extracted from the latest parts of projects' histories), we noticed a substantial drop in both F1 and ChrF scores.

We then evaluated the pre-trained versions of the models on the training parts of the collected per-project datasets (\ie the data that we use for fine-tuning the models). As it consists of projects' snapshots, it closely resembles the testing part of the \emph{Java-medium} dataset. \Cref{tab:time-investigation} shows the evaluation results on both training and testing parts of the collected data.

As we expected, the results we get on the training parts of the projects are lower, but still close to the ones reported in the papers. We suspect that a slight decrease in F1-score happens because the snapshots used here are different from the ones in the \emph{Java-medium} dataset, as we selected them specifically for the fine-tuning experiment (see \Cref{sec:dataset-mining}). 

However, when measured on the testing data from the collected projects, F1-score drops drastically. The major difference between training and testing data is the separation in time. For each project, the training data comes from its snapshot at some point of its history. Such snapshots contain both new and old methods. On the other hand, testing data includes the most recently developed methods in each project. It turns out that the prediction of names for new methods is more challenging for the models.

\begin{table}[]
\caption{Averaged F1-score for the pre-trained versions of studied models on training and testing parts of the datasets collected in this work.  }
\label{tab:time-investigation}
\centering
\begin{tabular}{ccc}
\toprule
\textbf{Model} & \textbf{Training data, F1} & \textbf{Testing data, F1} 
\\ \midrule
CodeTransformer & 48.5 & 40.9 \\ 
Code2Seq        & 47.1 & 34.8 \\ 
TreeLSTM        & 38.5 & 26.9 \\  \bottomrule 
\end{tabular}
\end{table}

The testing datasets that we collect for the fine-tuned models resemble the practical scenario of applying method name prediction models in practice. As it turns out to be more challenging for the models, we suggest that future research should take into account the time of creation for the code snippets used during evaluation. However, the observed phenomenon requires a more in-depth analysis.

\end{document}